# Metasurface-empowered freely-arrangeable multi-task diffractive neural networks with weighted training

Yudong Tian,[1, †] Haifeng Xu,[2, †] Yuqing Liu,[1] Xiangyu Zhao,[1] Jingzhu Shao,[1] Jierong Cheng,[2, *] and Chongzhao Wu[1, *]

[1] *Center for Biophotonics, Institute of Medical Robotics, School of Biomedical Engineering, Shanghai Jiao Tong University, Shanghai, China*

[2] *Institute of Modern Optics, Nankai University, Tianjin, China*

[†] *These authors contributed equally.*

*\* Corresponding author:* <u>chengjr@nankai.edu.cn (Jierong Cheng) and czwu@sjtu.edu.cn (Chongzhao Wu)</u>

Abstract: Recent advancements in optical computing have garnered considerable research interests owing to its energy-efficient operation and ultralow latency characteristics. As an emerging framework in this domain, diffractive deep neural networks ($D^2NNs$) integrate deep learning algorithms with optical diffraction principles to perform computational tasks at light speed without requiring additional energy consumption. Nevertheless, conventional $D^2NN$ architectures face functional limitations and are typically constrained to single-task operations or necessitating additional costs and structures for functional reconfiguration. Here, an arrangeable diffractive neural network (A-DNN) that achieves low-cost reconfiguration and high operational versatility by means of diffractive layer rearrangement is presented. Our architecture enables dynamic reordering of pre-trained diffractive layers to accommodate diverse computational tasks. Additionally, we implement a weighted multi-task loss function that allows precise adjustment of task-specific performances. The efficacy of the system is demonstrated by both numerical simulations and experimental validations of recognizing handwritten digits and fashions at terahertz frequencies. Our proposed architecture can greatly expand the flexibility of $D^2NNs$ at a low cost, providing a new approach for realizing high-speed, energy-efficient versatile artificial intelligence systems.

## 1. Introduction

Deep learning [1] has emerged as a transformative paradigm in artificial intelligence, demonstrating remarkable capabilities across diverse domains including computer vision [2–4], speech recognition [5,6], natural language processing [7], and autonomous driving [8,9]. However, as a data-driven algorithm, deep learning models require substantial computational resources and frequent data access, posing challenges for their deployment on conventional electronic computing systems. In traditional von Neumann architectures [10], the physical separation between memory and processing units introduces the computing performance bottleneck due to the speed mismatch between data retrieval and computation. Furthermore, the frequent shuttling of data between memory and processors incurs significant energy consumption, which is another block to the development of deep learning technologies.

In the recent decades, optical neural networks (ONNs) have emerged as a promising solution to address the limitations of conventional electronic computing systems, offering distinct advantages for deep learning in terms of their high parallel processing capability, lower energy consumption, and high-speed characteristics [11–16]. Recently, diffractive deep neural networks ($D^2NNs$), which use cascaded diffractive layers to physically implement deep learning with light, have emerged as a promising ONN architecture [17]. Similar to electronic neural networks, each pixel in the diffractive layers can be regarded as a neuron in the network, and the inter-layer connections are established through light diffraction. Compared to other optical architectures, one significant advantage of $D^2NNs$ is that each layer can be scaled up to millions of artificial neurons that are directly interconnected with neurons in adjacent layers. This ultrahigh density and parallelism endow the system with rapid and high-throughput computing capabilities, enabling a wide range of applications, including optical logic gate operation [18,19], multispectral snapshot imaging [20], quantitative phase imaging [21], visual recognition [22,23], and beam shaping [24,25], among others [26–30]. Metasurfaces, composed of sub-wavelength metallic/dielectric structures, are a novel type of artificial two-dimensional materials [31–34]. Unlike traditional 3D-printed diffractive optical elements (DOEs) that rely on phase accumulation during propagation, metasurfaces can precisely manipulate the incident electromagnetic wavefront through local and abrupt phase shifts caused by the interaction between electromagnetic waves and these meta-atom structures [35–37]. This unique characteristic enables metasurfaces to manipulate light at the wavelength scale, opening up new opportunities for the development of flexible and compact optical systems. The integration of metasurfaces into $D^2$NNs not only enables flexible manipulation of optical fields but also facilitates the development of miniaturized, multifunctional intelligent integrated devices [38]. Despite the remarkable achievements attained, the current $D^2$NN architectures still face significant scalability challenges due to their static, physically-fabricated nature. Once fabricated, these networks lack reconfigurability for new tasks. If other tasks need to be performed, the entire networks have to be retrained and manufactured, which consumes a lot of computing resources and reconstruction costs. Although some solutions, including programmable $D^2$NN [39], reconfigurable diffractive optoelectronic processor [40], pluggable diffractive neural network (P-DNN) [41], hardware-software co-design architecture [42], have been proposed to tackle these challenges, these methods typically require complex experimental setups and additional implementation costs, which may limit their practical applicability. Besides, research efforts have also explored the implementation of multi-task $D^2$NN by metasurface based optical multiplexing techniques [43,44]. However, these investigations have been exclusively confined to single-layer metasurface, limiting the potential for further enhancement of network performance.

In this work, we present an arrangeable diffractive neural network (A-DNN), which can switch to different tasks by altering the order of cascaded diffractive layers within the system. Moreover, we propose an efficient weighted training strategy for the design of A-DNN, which enables flexible control over the accuracy of different tasks. Finally, an A-DNN composed of two-layer cascaded phase-only metasurfaces [45] is designed for handwritten digit and fashion classification tasks [46,47], demonstrating the effectiveness of the proposed method. The results show that our multi-task A-DNN can achieve comparable accuracy for both tasks compared to the regular $D^2$NNs while significantly improving the hardware efficiency. Furthermore, as a general framework, A-DNN can be easily combined with other multiplexing techniques, which can greatly improve the scalability of the system. This flexible approach not only enables improved hardware efficiency and reconfigurability but also provides a new platform for developing integrated multifunctional artificial intelligence systems.

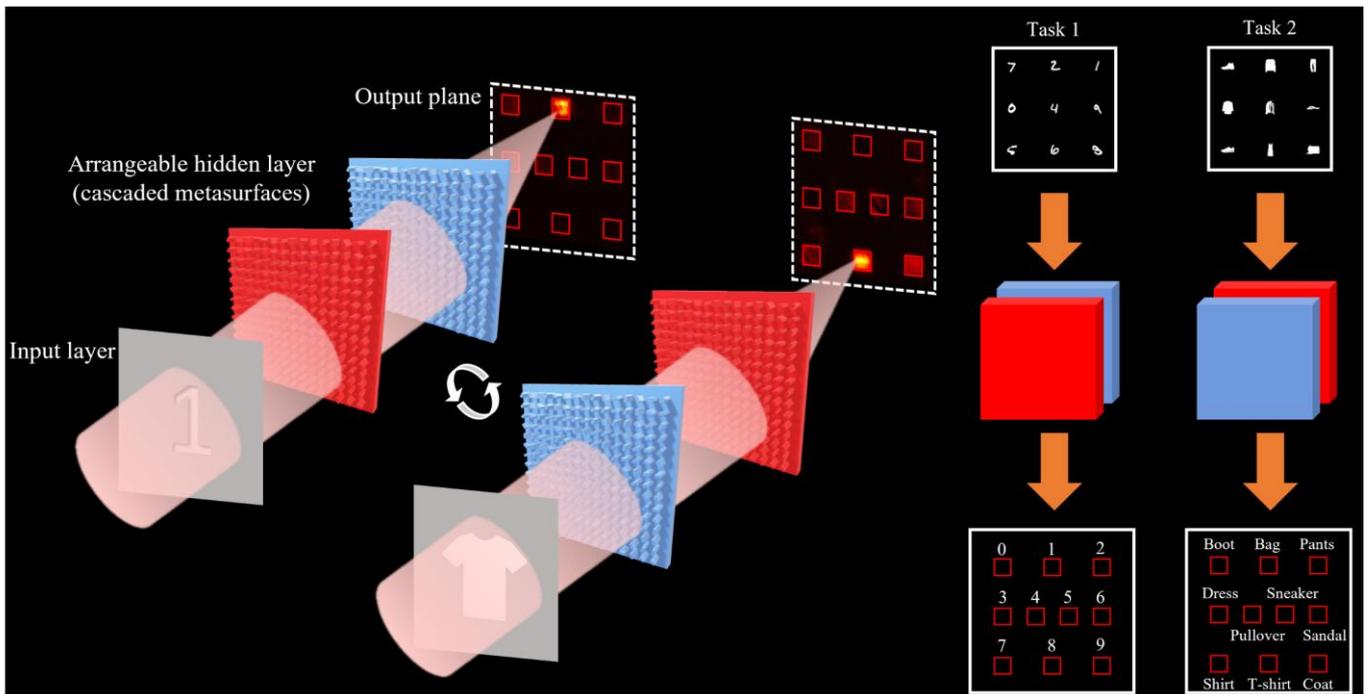

**Fig. 1. Framework of the arrangeable diffractive neural network (A-DNN) for multi-task learning.** The A-DNN comprises a series of cascaded diffractive layers, where the specific sequence of these layers determines the execution of distinct visual tasks. The recognition of handwritten digits and fashions can be achieved by altering the sequence of the two diffractive layers. A collimated light beam, encoded with amplitude information representing a specific object, serves as the input. After being modulated by the diffractive layers, the light is focused onto a designated detector region on the output plane for classification.

## 2. Method

Fig. 1 illustrates the framework of the proposed A-DNN. Specifically, the whole system mainly consists of three parts: the object to be classified as input, the cascaded metasurfaces containing different phase encodings as the hidden layer, and the discretized detection plane as the output layer. By altering the sequence of the cascaded metasurfaces, the phase encoding of the hidden layer can be changed to adapt to different visual tasks. The input plane functions as a mask containing the pattern to be identified. When incident plane wave propagates through the input layer, the amplitude information of the corresponding object is obtained. Subsequently, the diffracted light is focused onto the output plane after being modulated by two metasurface layers. The output plane is partitioned into multiple discrete small detection regions, where the number of regions corresponds to the number of categories in the target dataset, with each region representing a specific class. The predicted class of the network is generated by selecting the index of the highest probability output, which corresponds to the detection region with the maximum energy value.

Like traditional D²NNs, the training of A-DNN is also based on the optical diffraction theory [17,48]. According to the Rayleigh-Sommerfeld diffraction equation, the wave propagation between adjacent diffractive layers can be treated as the secondary sources composed of the following optical mode:

$$w_i^l(x,y,z) = \frac{z - z_i}{r_i^2} \left( \frac{1}{2\pi r_i} + \frac{1}{j\lambda} \right) \times exp\left( \frac{j\, 2\pi r_i}{\lambda} \right) \tag{1}$$

where $(x_i, y_i, z_i)$ represents the location of the $i^{th}$ neuron on the $l^{th}$ layer, $\lambda$ is the wavelength, $j = \sqrt{-1}$, and $r_i$ is the distance of this diffractive process, which equals $\sqrt{(x-x_i)^2 + (y-y_i)^2 + (z-z_i)^2}$. The amplitude and phase of the secondary wave are determined by the input wave and the transmission coefficient $t$ of the neuron, which can be expressed as:

$$n_i^l(x,y,z) = w_i^l(x,y,z) \cdot t_i^l(x_i, y_i, z_i) \cdot \sum_k n_k^{l-1}(x_i, y_i, z_i) \tag{2}$$

where $\sum_k n_k^{l-1}(x_i, y_i, z_i)$ represents the incident waves to the $i^{th}$ neuron of layer $l$, and $t_i^l(x_i, y_i, z_i) = a_i^l(x_i, y_i, z_i) \cdot exp(\varphi_i^l(x_i, y_i, z_i))$ is composed of amplitude and phase term. In this paper, since we use phase-only encoding, the amplitude term $a_i^l(x_i, y_i, z_i)$ is equal to a constant, and only the phase value is used as a learnable parameter to construct the network.

Let $U^l(x_l, y_l)$ denote the light field distribution as the incident light propagates to the $l^{th}$ layer. Given that the diffractive neural network consists of $M$ layers, the light field intensity detected at the output plane can be expressed as:

$$I^{M+1} = |U^{M+1}(x_{M+1}, y_{M+1})|^2 \tag{3}$$

Subsequently, by selecting an appropriate loss function to quantify the error, the network parameters can be updated using the backpropagation algorithm based on the gradient descent method. Here, we define the loss function $L$ between the light intensity at the output plane $I^{M+1}$ and the target $Y^{M+1}$ as follows:

$$\mathcal{L}(\varphi^1, \varphi^2, \dots, \varphi^M) = F(I^{M+1}, Y^{M+1}) \tag{4}$$

where $\varphi^l, l = 1,2,\cdots, M$ represents the phase of $l^{th}$ layer and $F$ represents the error function between $I^{M+1}$ and $Y^{M+1}$, such as the mean square error (MSE). Ultimately, the training process of a neural network aims to optimize the following problem:

$$\min_{[\varphi^1, \varphi^2, \cdots, \varphi^M]} \mathcal{L}(\varphi^1, \varphi^2, \cdots, \varphi^M), s.t.\, 0 \leq \varphi_i^l(x_i, y_i, z_i) \leq 2\pi \tag{5}$$

Unlike other multi-task D²NN frameworks [41,43], A-DNN requires only a single backpropagation step to optimize multiple tasks, which is achieved by computing a weighted summation of the loss functions for different tasks. More specifically, assuming that an A-DNN is designed to handle $N$ tasks $\{[X^1, Y^1], [X^2, Y^2], \cdots [X^N, Y^N]\}$, where $X^i$ represents the input space of task $i$, and $Y^i$ represents corresponding solution space, the loss function of the A-DNN is defined as follows:

$$\min_{[\varphi^1, \varphi^2, \cdots, \varphi^M]} \sum_{n=1}^{N} \beta^n \mathcal{L}^n(\theta^n; X^n, Y^n) \tag{6}$$

where $\mathcal{L}^n$ is the loss function of the $n^{th}$ task, $\theta^n$ represents the corresponding order of diffractive layers, and $\beta^n$ is the weight assigned to the $n^{th}$ task. Finally, the optimization of multiple tasks can be accomplished by simply performing one backpropagation on this weighted function. Meanwhile, the accuracy of different tasks can be controlled by assigning weights of varying magnitudes.

In this work, we designed and demonstrated the multi-task A-DNN with a two-task A-DNN architecture shown in Fig. 1. This network consists of two metasurfaces, which can generate two configurations respectively designed for the classification tasks of handwritten digits and fashions. The sequence of the diffractive layers is defined as $\theta^1 = \{\varphi^1, \varphi^2\}$ and $\theta^2 = \{\varphi^2, \varphi^1\}$, and the multi-task loss function can be written as:

$$\mathcal{L}_{multi} = \mathcal{L}(\theta^1; X^1, Y^1) + \beta \mathcal{L}(\theta^2; X^2, Y^2) \tag{7}$$

where we simply set $\beta^1 = 1$ and $\beta^2 = \beta$. The complete training process of the two-task A-DNN is illustrated in Fig. 2.

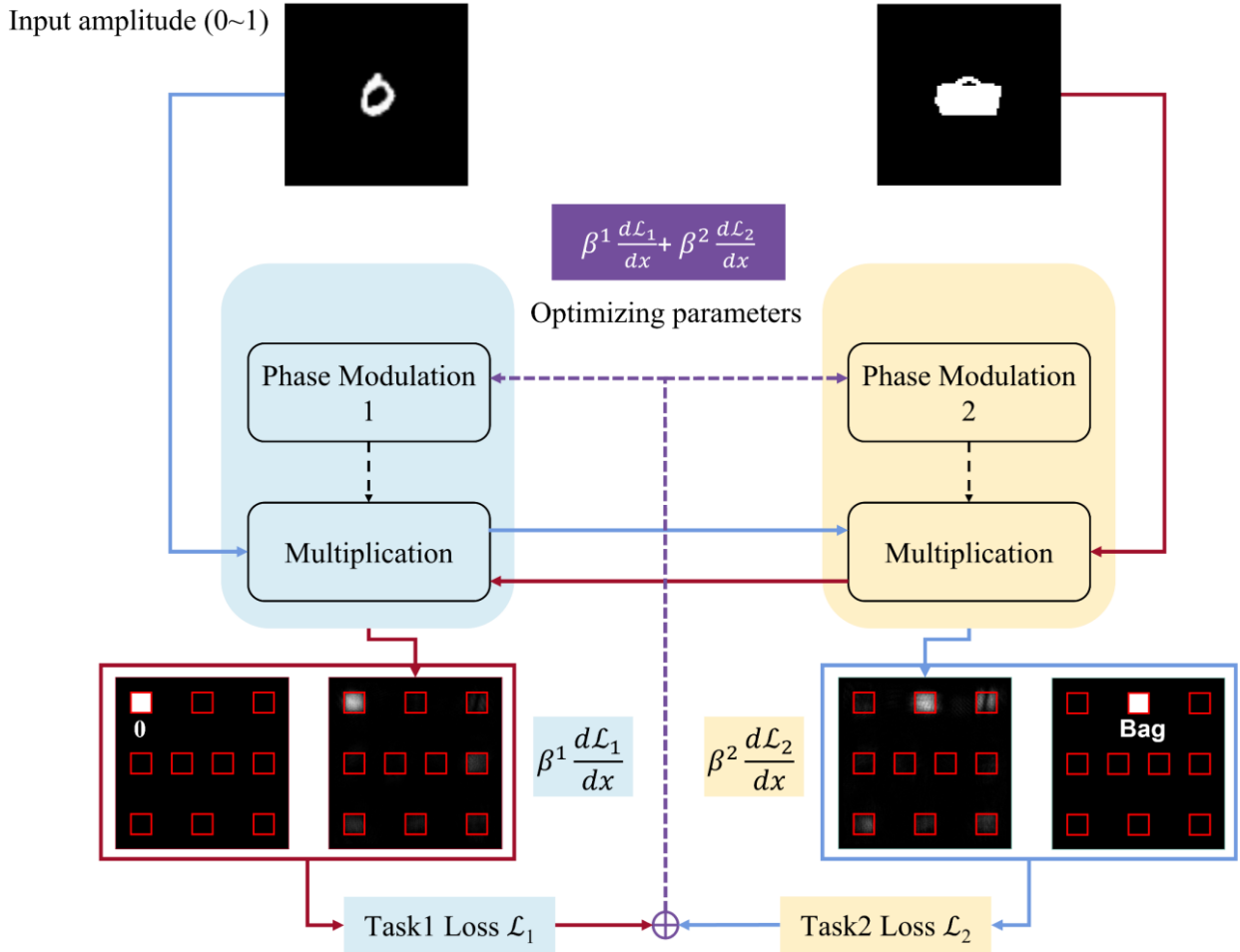

**Fig. 2. Flowchart of weighted training process for the two-task A-DNN.** The forward propagation of the input light wave is computed using the angular spectrum method. As the light wave propagates through each diffractive layer, the complex field is modulated by multiplying the corresponding phase profile. Subsequently, losses of different tasks are combined through weighted summation, enabling parameter updates via a single backpropagation step. This approach allows for the assignment of distinct attention weights to different tasks, facilitating task-specific optimization.

## 3. Results

### 3.1 Evaluation of weighted training for biased multi-task learning

In multi-task learning scenarios, the adjustment of task weights or importance levels is often required to accommodate specific application requirements. In this section, we demonstrate the effectiveness of our proposed weighted training strategy in addressing such biased multi-task learning problems. Specifically, we maintain a fixed weight of 1 for the handwritten digit classification task, while adjusting the weight parameter β for the fashion classification task to regulate the relative performance of these two tasks (see Eq. 7). Fig. 3 depicts the training and testing performance metrics of A-DNN across different β configurations. It can be observed that as β increases (or decreases), the classification performance of A-DNN on the

FashionMNIST (or MNIST) dataset correspondingly improves (or declines), demonstrating the effectiveness of weighted training approach in biased multi-task learning. Moreover, it can be observed that moderately increasing the task weight for FashionMNIST leads to an overall improvement in the network's accuracy (Fig. 3d). This suggests that appropriately assigning higher weights to more challenging tasks can enhance the overall performance of the model.

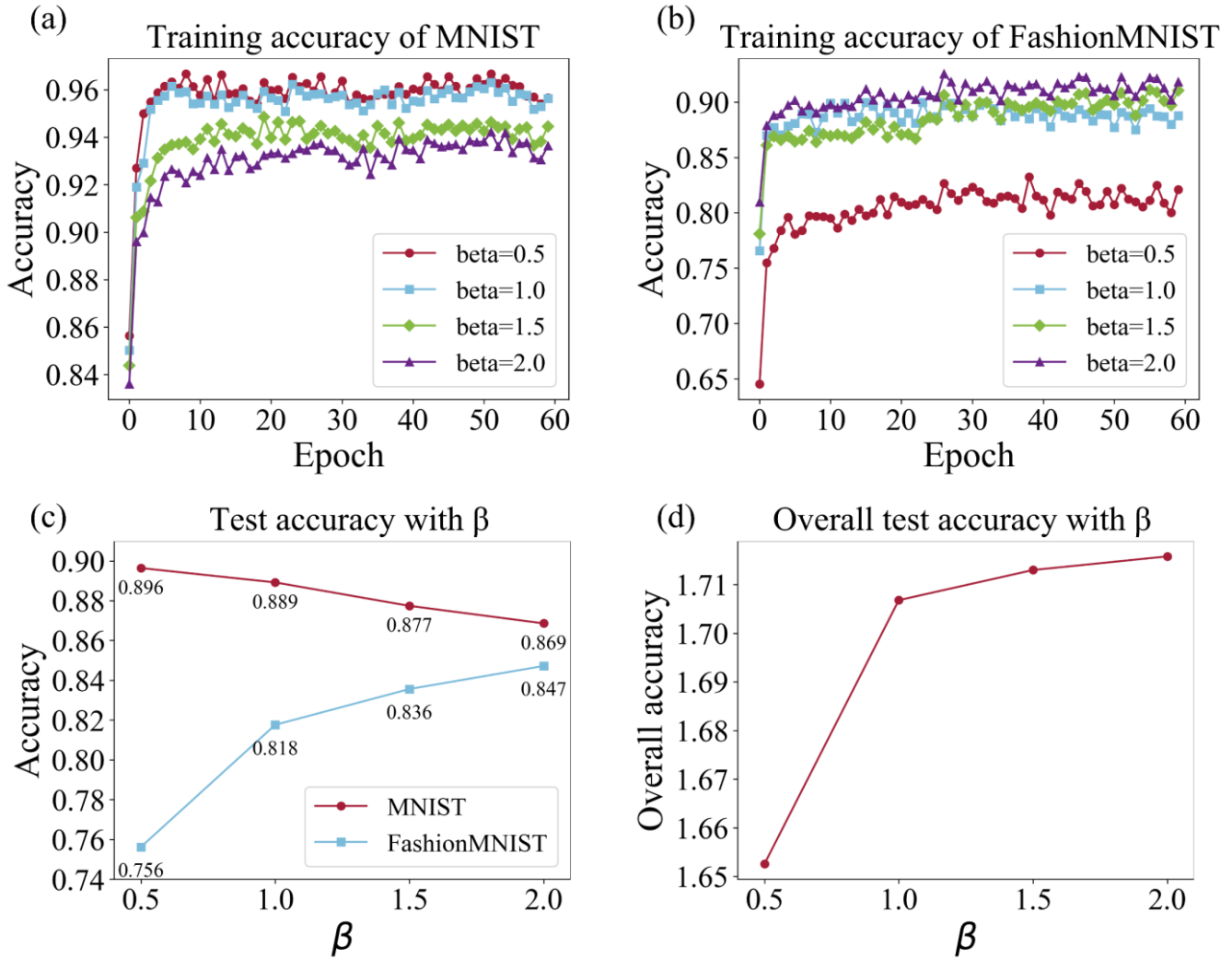

**Fig. 3. Evaluation of weighted training for task-specific performance adjustment.** (a) Accuracy convergence curves for the MNIST classification task under varying values of β during the training phase; (b) Accuracy convergence curves for the FashionMNIST classification task under varying values of β during the training phase; (c) Test accuracy with different weight parameter β. As β increases (or decreases), the final performance of the multi-task A-DNN shifts toward favoring the FashionMNIST (or MNIST) task; (d) Overall test accuracy of A-DNN with different weight parameter β.

### 3.2 Comparison of A-DNN performance against the regular D²NN architecture

Next, we compare the performance of our A-DNN architecture against a regular $D^2NN$ structure for handwritten digit and fashion classification tasks. In this comparison, we employ the A-DNN model described in Section 3.1, trained with β = 1, which incorporates a total of $N_a$ = 80000 diffractive features. And for original $D^2NN$ structure, we train separate two-layer network for handwritten digit and fashion classification respectively, with each layer sized at 200 × 200 pixels, yielding a total of N = 2×$N_b$ trainable diffractive features. The original $D^2NN$ is trained using the same hyperparameters, image datasets, and the loss functions as used for A-DNN. Upon completion of the training, both the A-DNN and the $D^2NN$ are evaluated using the identical test dataset to ensure a fair comparison of their performance.

Fig. 4 and Table 1 present the results of the comparison experiment. We can see that, whether on the training dataset or the test dataset, the A-DNN achieves performance comparable to that of the $D^2NN$ across both tasks, while utilizing only half the number of diffractive units employed by the $D^2NN$, resulting in a 50% improvement in hardware efficiency. Simultaneously, the A-DNN is capable of optimizing all tasks through a single training session, thereby saving nearly half of the training time compared to the regular $D^2NN$. Moreover, the advantages of the A-DNN become increasingly pronounced as the number of tasks and diffractive layers grow. In a three-task A-DNN composed of four diffractive layers (see section S1 in supplementary material), the improvement in hardware efficiency further escalates to 66.7%, while the training time is reduced to merely one-third of that required by regular $D^2$NNs.

Fig. S2 presents the results of using two single-layer $D^2$NNs to handle handwritten digit and fashion classification tasks separately. It is evident that compared to the networks with cascaded diffractive layers, the accuracy of single-layer $D^2$NNs significantly decreases (by approximately 20%), highlighting the importance of cascaded diffractive layers in enhancing system performance.

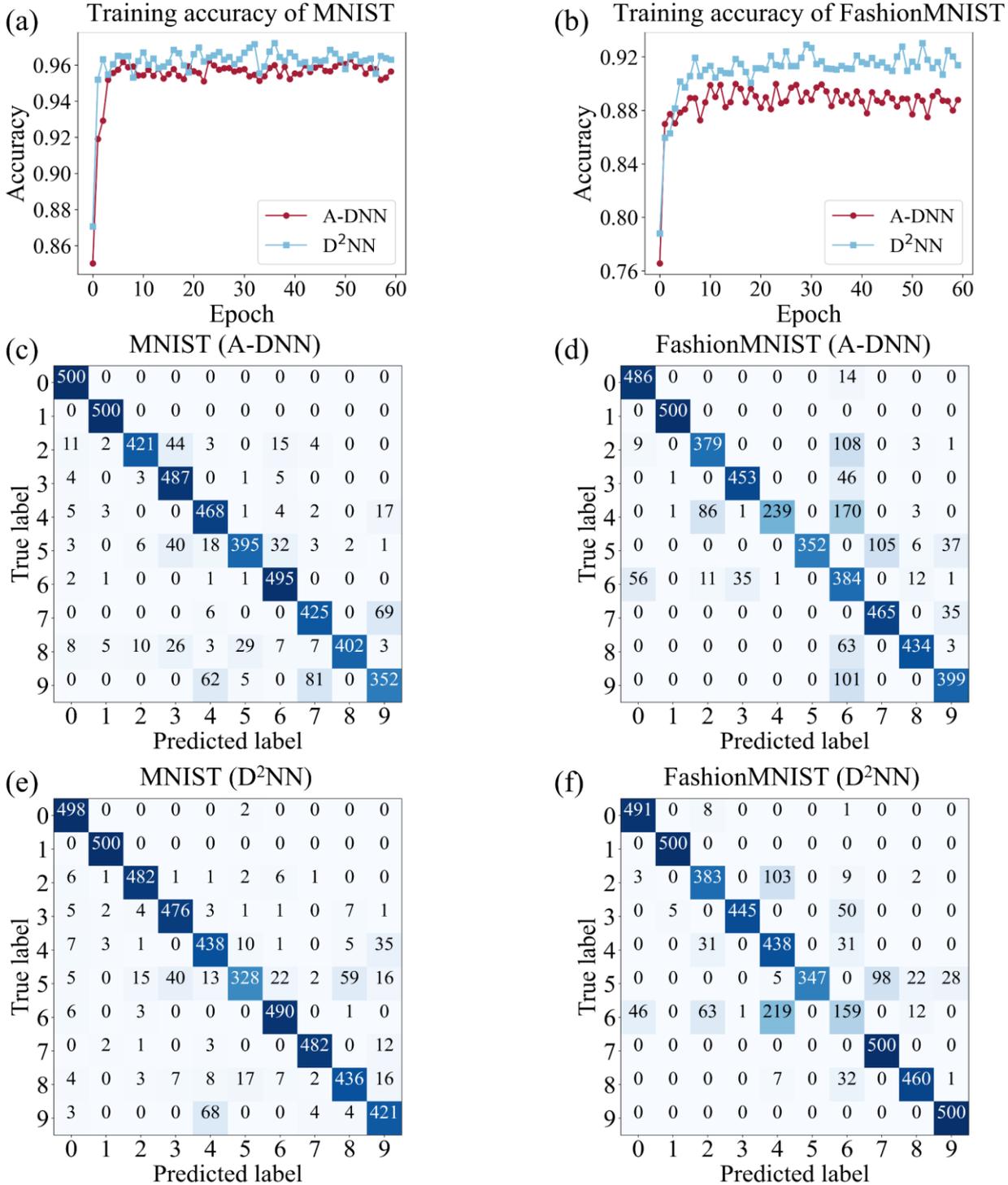

**Fig. 4. Performance comparison between the A-DNN and regular $D^2$NN architecture.** (a)(b) Accuracy convergence curves of the A-DNN and regular $D^2$NN on the MNIST and FashionMNIST tasks, demonstrating only marginal performance differences between the two architectures; (c)(d) Confusion matrices of the A-DNN on the MNIST and FashionMNIST test datasets; (e)(f) Confusion matrices of the $D^2$NN on the MNIST and FashionMNIST test datasets.

Table 1. Hardware cost and performance comparison between the regular $D^2$NN and multi-task A-DNN architecture.

|  | Regular $D^2$NN architecture | | A-DNN architecture | |
| --- | --- | --- | --- | --- |
| Dataset | MNIST | FashionMNIST | MNIST | FashionMNIST |
| Diffractive layer cost | 2×200×200 | 2×200×200 | 2×200×200 | |

| Training accuracy | 0.971 | 0.918 | 0.964 | 0.891 |
| Test accuracy | 0.910 | 0.844 | 0.889 | 0.818 |

### 3.3 Robustness test of A-DNN

To evaluate the robustness of the proposed architecture, we conduct a theoretical analysis to investigate the impact of alignment errors and phase noise in the diffractive layers on the model's performance. Specifically, we simulate displacement errors by shifting the position of the first diffractive layer along the $x$, $y$, and $z$ axes (Fig. 5a, b), and we model potential modulation phase errors arising from fabrication imperfections by adding random Gaussian noise to the modulation phase of the diffractive layers. Fig. 5c illustrates the variation in the classification accuracy of the A-DNN under Gaussian noise with a mean of 0 and different standard deviation ($\sigma$). It can be observed that the A-DNN maintains robust performance for both tasks (accuracy drop is less than 5%) even under Gaussian noise with $\sigma = 0.3\pi$, demonstrating the architecture's strong resilience to modulation phase errors introduced during fabrication. Fig. 5d-f illustrate the classification accuracy of the A-DNN under different displacements of the diffractive layers, where $x$ and $y$ represent the side lengths of the diffractive layers, $z$ denotes the distance between the diffractive layers, and $\Delta x$, $\Delta y$, and $\Delta z$ indicate displacements in the corresponding directions. The results indicate that the A-DNN maintains robust performance on both tasks when the $z$-axis displacement is less than 5% and the $x$- and $y$- axis displacements are less than 2%. Additionally, it is observed that the A-DNN exhibits greater resilience to displacement errors in the fashion classification tasks. This may be attributed to the fact that, in handwritten digit classification, the displaced diffractive layer is positioned at the first layer, primarily serving a feature extraction role. Noise introduced during this process can affect the input to subsequent layers and be progressively amplified, thereby impacting the overall model performance. In contrast, for fashion classification, this diffractive layer is located at the last layer, making it less directly susceptible to noise induced by displacement errors.

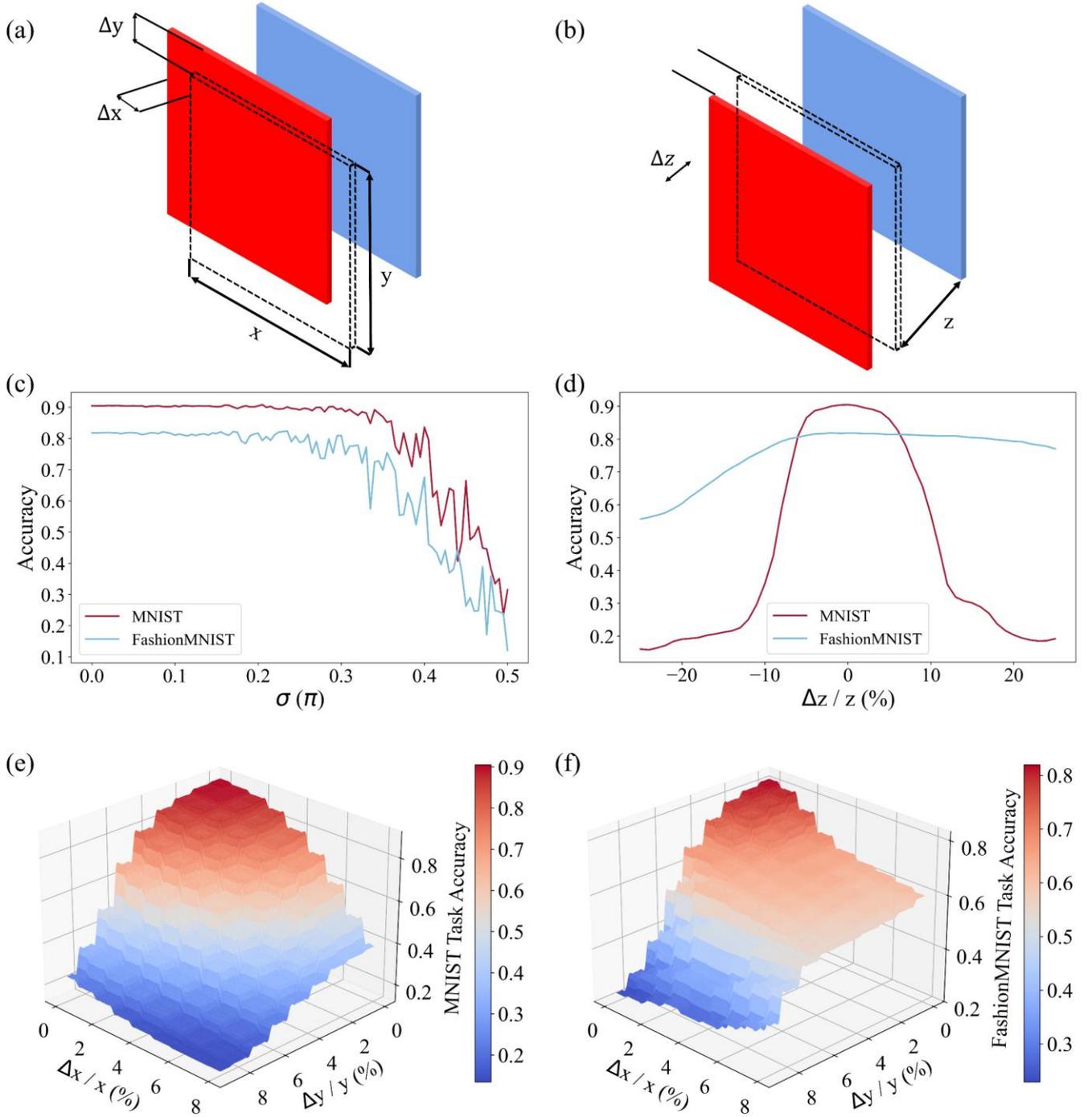

**Fig. 5. Robustness evaluation of the proposed multi-task A-DNN against system noise and alignment errors.** (a, b) Schematic diagram of network alignment errors: *x*, *y*, and *z* represent the dimensions of the diffraction layer along the *x*-, *y*-, and *z*-axis, respectively, and Δ*x*, Δ*y*, and Δ*z* denote the displacements along the *x*-, *y*-, and *z*-axis, respectively. (c) Prediction accuracy under varying levels of Gaussian noise; (d) Analysis of alignment errors in the *z*-direction for the A-DNN; (e, f) Analysis of alignment errors in the *x*- and *y*-directions for the A-DNN: (e) Accuracy of the MNIST task under different *x*-axis and *y*-axis displacements; (f) Accuracy of the FashionMNIST task under different *x*-axis and *y*-axis displacements.

### 3.4 Experimental demonstration of cascaded metasurfaces empowered A-DNN

As a proof-of-concept, we theoretically design and experimentally validate a two-layer cascaded metasurface-based A-DNN architecture tailored for handwritten digit and fashion classification tasks. The A-DNN is trained at 0.291 THz ($\lambda = 1030$ μm) with $2 \times 80 \times 80$ neurons (12800 in total), and the layer-to-layer axial distance is fixed at 15 mm. The chosen training datasets are subsets of the MNIST and FashionMNIST, containing digits 1-5 and five fashion items—T-shirts, trousers, dresses, bags, and sneakers, respectively (Fig. S5). The building block of the metasurfaces is chosen as a silicon rectangular nanofin on top of the silicon substrate (Fig. 6a). By changing the rotation angle of the meta-atoms, the Pancharatnam-Berry (PB) phase modulation mechanism provides full phase control from 0 to $2\pi$ for circularly polarized light [49]. The metasurface design is implemented using the finite-difference time-domain (FDTD) method.

The height of the nanofins is fixed at 500 μm, and the periods are set to 515 μm in both the *x* and *y* directions. Parameter sweeps of the nanofin's length and width are conducted in 5 μm increments, ranging from 50 μm to 500 μm. Ultimately, a nanofin with a length (L) of 90 μm and a width (W) of 465 μm, which achieves a remarkable polarization conversion efficiency of 91.2%, is selected to construct the metasurfaces. Fig. 6b presents the simulated transmission coefficients for both circularly cross-polarized and co-polarized components. When the incident light is set as left-handed circularly polarized (LCP), the relationship between transmittance, additional phase, and the rotation angle of the optimized nanofin is shown in Fig. 6d, demonstrating that the additional phase can span the full range from 0 to $2\pi$. Fig. 6e exhibits the optimized phase modulation maps of A-DNN and the optical images of the corresponding fabricated metasurfaces. The effective area of the metasurface is $4.12 \times 4.12$ cm$^2$ and contains $80 \times 80$ nanofins. The experimental system for demonstrating the performance of metasurface-based two-task A-DNN is shown in Fig. 6f. Further details of the experimental setup can be found in the Experimental Section.

During the training stage, the A-DNN converges on both tasks after 5 epochs of iterative training, achieving simulated test accuracies of 93.1% on MNIST and 87.2% on FashionMNIST, respectively (Fig. 7a, Fig. 8a). We use 5000 handwritten digit and fashion images as the test datasets (500 images per classification category) and fabricate 40 samples for experimental validation of the metasurface-empowered A-DNN. The test results are presented in the confusion matrices detailing the correctly identified and misidentified instances for both simulation and experimental data, as shown in Fig. 7b-c and Fig. 8b-c. According to the statistical results, the experimental classification accuracies for handwritten digits and fashions are 75% and 70%, respectively. Fig. 7d-e and 8d-e present the input patterns and their corresponding angular spectrum diffraction results, respectively. The fabricated masks and measured diffractive field distributions are shown in Fig. 7f–g and Fig. 8f-g. The simulated and measured energy distributions are illustrated in Fig. 7h and Fig. 8h, demonstrating the designed A-DNN's ability to simultaneously recognize digits and fashions. Additional results from simulations and experiments can be found in the section S4 of supplementary materials.

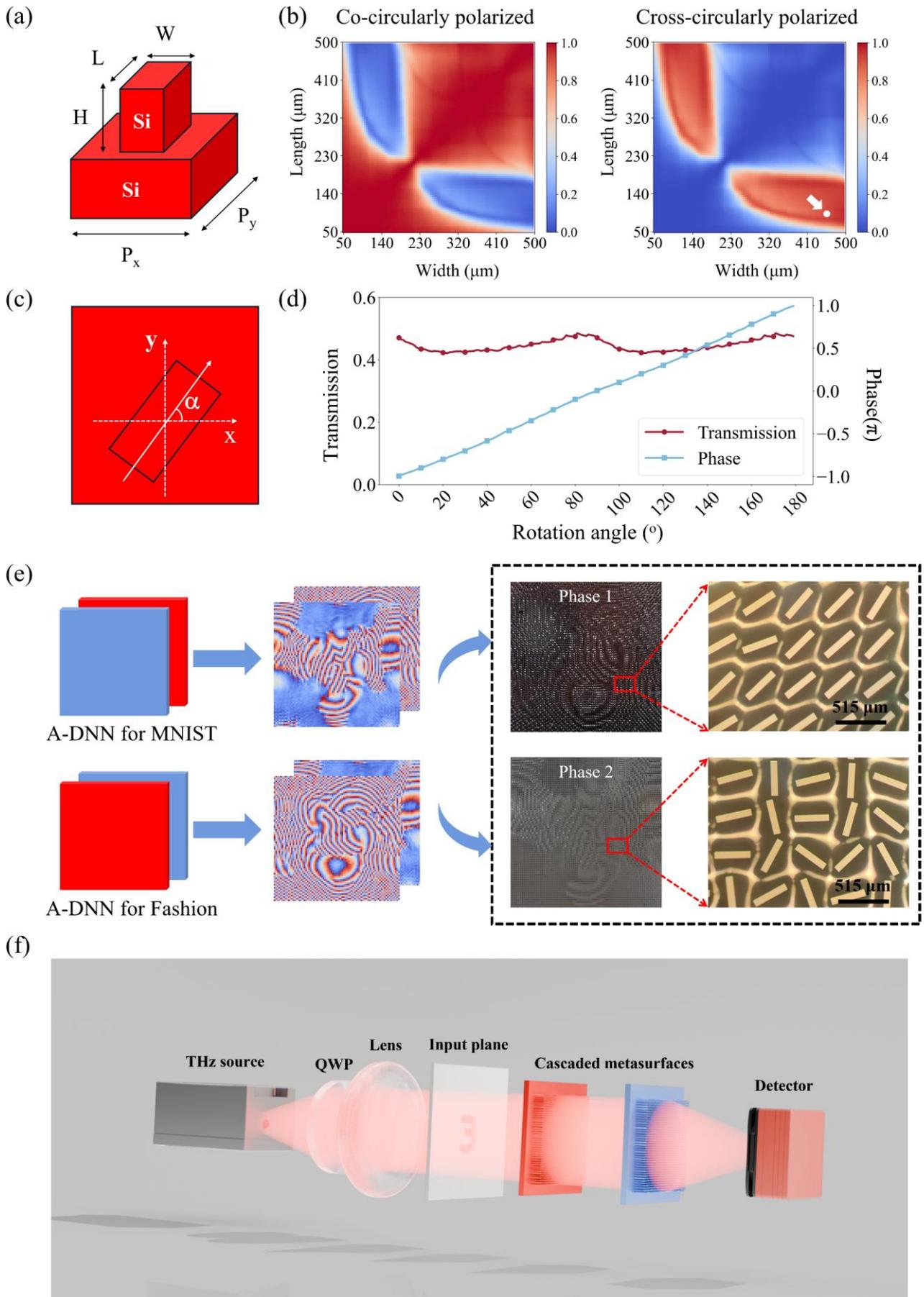

**Fig. 6. Design of the metasurface-empowered A-DNN and experimental setup.** (a) Schematic of a high-resistivity silicon meta-atom fabricated on a high-resistivity silicon substrate. Here, $P_x$ and $P_y$ represent the periods in the *x*- and *y*- directions, H denotes the height of the meta-atom, and L and W correspond to the length and width of the individual meta-atom, respectively; (b) Amplitude map of the circular transmission coefficient in cross- and co-polarization under meta-atom with different geometric dimensions, resulting in the structural parameters of L = 90 μm and W = 465 μm; (c) Schematic for the deflection angle of the meta-atom; (d) Relationship between the additional phase, transmission, and the rotation angle of the meta-atom; (e) Optimized phase distribution of A-DNN diffractive layers and the optical images of the fabricated metasurfaces; (f) Experimental setup for the characterization of metasurface-empowered A-DNN.

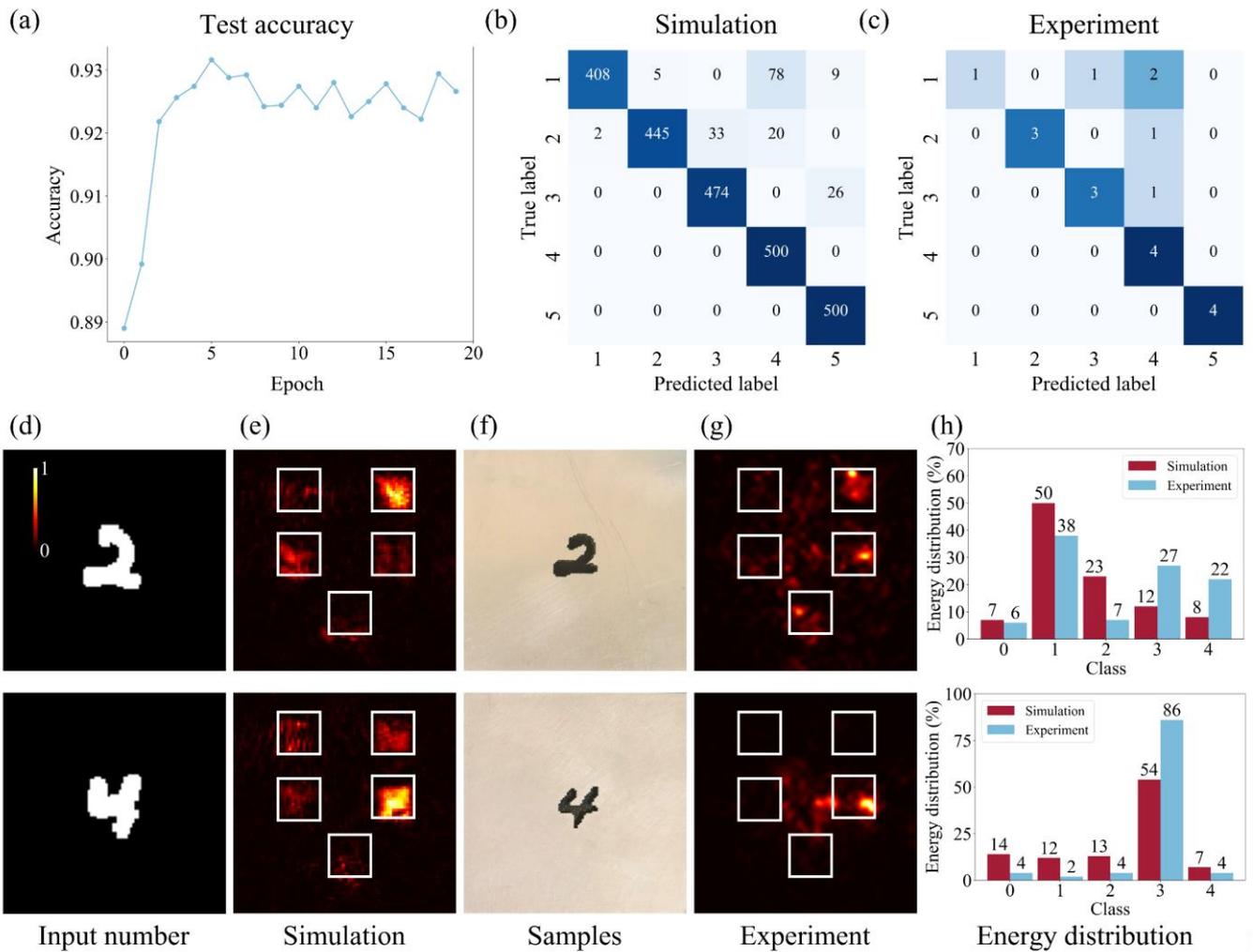

**Fig. 7. Simulation and experimental results of A-DNN for handwritten digits classification.** (a) Test accuracy convergence curve during training stage: the accuracy stabilizes at 93% after 20 epochs; (b)(c) Confusion matrices for handwritten digit classification from simulation and experimental results, respectively; (d) Amplitude images of input numbers; (e) Normalized energy distribution on the output plane obtained from simulations; (f) Optical images of the fabricated handwritten digit samples; (g) Normalized energy distribution on the output plane obtained from experiments; (h) Energy distribution percentage across different detection regions in both simulations and experiments.

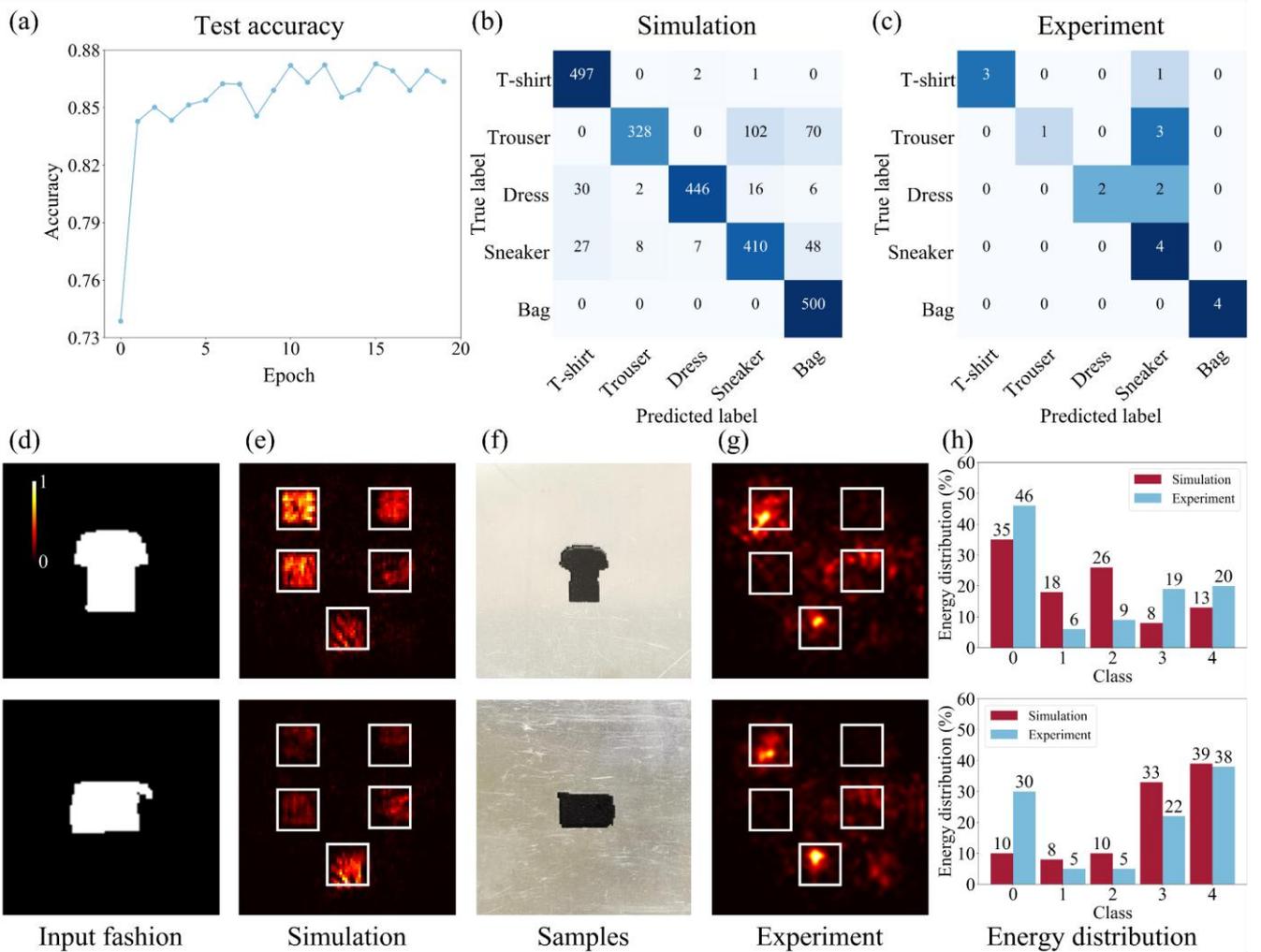

**Fig. 8. Simulation and experimental results of A-DNN for fashions classification.** (a) Test accuracy convergence curve during training stage: the accuracy stabilizes at 87% after 20 epochs; (b)(c) Confusion matrices for fashion classification from simulation and experimental results, respectively; (d) Amplitude images of input fashions; (e) Normalized energy distribution on the output plane obtained from simulations; (f) Optical images of the fabricated fashion samples; (g) Normalized energy distribution on the output plane obtained from experiments; (h) Energy distribution percentage across different detection regions in both simulations and experiments.

## 4. Discussion and conclusion

In this study, we present the multi-task arrangeable diffractive neural network (A-DNN) to address the lack of reconfigurability in traditional D$^2$NN architectures. By altering the sequence of the cascaded diffractive layers within the network, the A-DNN can be reconfigured to handle multiple distinct visual tasks. Furthermore, a weighted training strategy is proposed for training the A-DNN, which enables flexible control over the performance of different tasks while significantly reducing training time. To verify its feasibility, we design an A-DNN for handwritten digit and fashion recognition based on angular spectrum diffraction and implement it using cascaded metasurfaces. Although the experimental test accuracy is slightly lower than that achieved in numerical simulations, which may be attributed to errors introduced during the forward design process of the metasurfaces (see Fig. S8), the results collectively demonstrate the capability of our fabricated A-DNN in multi-task recognition. And these forward design errors can be effectively mitigated by constructing each pixel in the network through the repetitive arrangement of nano-fins with identical azimuth angles (also see Section S6). In the future, the performance of the metasurface can be enhanced through more advanced inverse design methods, and other optical multiplexing techniques can be further integrated into the A-DNN framework, which could significantly improve the system's classification performance and enhance the flexibility of the neural network. In summary, this versatile approach not only offers a convenient and powerful solution to address the limited functionality of traditional D$^2$NNs, but also opens up more possibilities for future development of high-speed parallel computing and multi-functional artificial intelligence systems.

## 5. Experimental Section

Training: The design of all models in this work is performed using Python (v3.12.4) and PyTorch (v 2.1.0) on a Windows 11 operating system (Microsoft) with AMD Ryzen 9 7950X 16-Core central processing unit, 128 GB of RAM, and a GeForce RTX 3090 graphical processing unit. The MNIST and FashionMNIST datasets are used for training with a training batch size of 32 and a learning rate of 0.5. The mean square error is employed as the loss function and the Adam optimizer is used to update the phase value of each layer in the network. For the ten-class classification tasks, each diffractive layer consists of 200×200 pixels with a pixel size of 515 μm, and the distance between diffractive layers is 10 cm. For the five-class classification tasks, each diffractive layer comprises 80×80 pixels with a pixel size of 515 μm, and the distance between diffractive layers is 1.5 cm.

Fabrication: The metasurfaces are fabricated on high-resistivity silicon wafer by contact photolithography and reactive ion beam etching. The thickness of the silicon wafer is 1 mm. The masks for the experiment are fabricated on 1mm-thick aluminum plates by laser cutting technology.

Experimental Setup: To detect the electric-field intensity distributions after the cascaded metasurfaces, we establish the experiment system shown in Fig. 6f. The THz source is an IMPATT diode emitting at 0.291 THz. A quarter-wave plate (QWP) is inserted to adjust the polarization of the beam. The radiation is then collimated by a TPX lens before illuminating the sample. The light field distribution at the output plane is recorded by a wiredSense pyroelectric detector with a scanning step size of 1 mm.

## Acknowledgements

This research was sponsored by the National Natural Science Foundation of China (62375170, 62235004, 62175118), the Shanghai Jiao Tong University (YG2024QNA51) and the Science and Technology Commission of Shanghai Municipality (20DZ2220400).

## Conflict of Interest

The authors declare no conflict of interest.

## Data Availability Statement

All the data used in this work are publicly available in PyTorch. The code is available at: Atrf/Arrangeable-multi-task-diffractive-neural-network: Code for Freely Arrangeable diffractive neural network.